\documentclass[a4paper,11pt]{article}
\pdfoutput=1 

\usepackage{jheppub} 
                     
\usepackage{jheppub}
\usepackage{mathtools,amssymb,amsmath,braket,mathrsfs,tikz,xspace,xcolor,float,color,siunitx,comment,afterpage,multirow,array,hhline,amsthm,graphbox,url}

\usepackage[T1]{fontenc} 
\usepackage{graphicx}
\usepackage{subfigure}
\usepackage{slashed}
\usepackage{indentfirst}
\usepackage{amsmath}
\usepackage{bm}

\usepackage{booktabs}
\usepackage{multirow}

\begin{document}

\title{\boldmath Testing type II seesaw leptogenesis at the LHC }



\author{Chengcheng Han,}

\author{Zhanhong Lei,}

\author[1]{Weihao Liao\note{Corresponding author.}}

\vspace{0.5cm}

\affiliation{School of Physics, Sun Yat-Sen University, Guangzhou 510275, P. R. China}

\emailAdd{hanchch@mail.sysu.edu.cn}
\emailAdd{leizhh3@mail2.sysu.edu.cn}
\emailAdd{liaowh9@mail2.sysu.edu.cn}

\abstract{
Type II seesaw leptogenesis simultaneously explains the origin of neutrino masses, the baryon asymmetry of our universe, and the inflation. The Large Hadron Collider(LHC) provides an opportunity to directly test  type II seesaw leptogenesis by looking for the predicted triplet Higgs. In this paper, we perform an analysis of the detection prospect for the triplet Higgs at the LHC through the multi-electron channels. We find that due to the contribution of $pp\to H^{\pm \pm }H^{\mp }$ process,  the sensitivity of multi-electron channels searching for the doubly-charged Higgs pair production can be improved. We also investigate the $3e+ {E}^{\rm miss}_{\rm T}$ signals to probe the $pp\to H^{\pm \pm }H^{\mp }$ production and find that the future high luminosity LHC could probe a triplet Higgs around 1.2 TeV at $2\sigma$ level.
}

\maketitle
\flushbottom

\section{Introduction}
\label{sec:intro}

One of the unresolved issues in modern physics is the origin of the neutrino mass. In the standard model(SM) neutrinos are massless, but the observation of the neutrino oscillation indicates that neutrinos have tiny masses, which requires the extension of the SM. The most popular ideas for generating neutrino masses are the so-called seesaw mechanisms, which can be classified into three types. The type I/III seesaw introduces additional three(at least two) singlet/triplet fermions~\cite{Minkowski:1977sc, Yanagida:1979as, Glashow:1979nm, Gell-Mann:1979vob, Foot:1988aq}, while the type II seesaw only includes an additional triplet scalar which provides a minimal framework to explain the origin of neutrino masses~\cite{Magg:1980ut, Cheng:1980qt, Lazarides:1980nt, Mohapatra:1980yp, PhysRevD.22.2227, PhysRevD.25.774}. In the model of type II seesaw, the triplet Higgs can directly couple to the lepton sectors, and if the neutral component of the triplet Higgs gets a vev, the Majorana mass of the neutrinos can be generated.  

 Interestingly, the type II seesaw could also provide a feasible leptogenesis if it also plays the role of the inflaton, as pointed out by a recent study \cite{Barrie:2021mwi, Barrie:2022cub}. Therefore, this simple model could explain three important problems at the same time: the origin of the neutrino masses, the baryon asymmetry of our universe, and inflation. Comparing with the leptogenesis from type-I seesaw which generally requires a high scale right-handed neutrino~\cite{Giudice:2003jh}, the type II seesaw leptogenesis allows the triplet Higgs to be as light as TeV scale which could be directly probed by the Large Hadron Collider(LHC). Indeed LHC already performs some surveys and currently sets a limit of around a few hundred GeV for the doubly charged Higgs contained in the triplet Higgs depending on its decay products \cite{CMS:2017pet,CMS:2017fhs,ATLAS:2017xqs,ATLAS:2018ceg,ATLAS:2021jol}. The decay of the doubly charged Higgs is sensitive to the vacuum expectation value of the triplet Higgs. For a large vev $v_\Delta \gtrsim 0.1$ MeV, it mainly decays into two gauge bosons,  otherwise, it would decay into dileptons \cite{PhysRevD.78.015018}. However, if the baryon asymmetry is generated by the type II seesaw, to avoid the lepton number to be washed out, the vev of the triplet Higgs $v_\Delta < 1$ keV is preferred. Therefore, looking for the triplet Higgs through the leptonic channel would provide a visible way to test the type II seesaw leptogenesis. 
In this paper, we investigate the detection capability of the triplet Higgs in future large hadron colliders. Previous studies on this aspect have been investigated in numerous works including, for example, Refs. \cite{FileviezPerez:2008wbg,Alloul:2013raa, Han:2015sca, Han:2015hba, Du:2018eaw, BhupalDev:2018tox,deMelo:2019asm, Primulando:2019evb,Padhan:2019jlc, Ashanujjaman:2021txz, Mandal:2022zmy, Duarte:2022xpm}. The test of the type II seesaw leptogenesis from lepton flavor violation can be also found in \cite{Barrie:2022ake}.


In the model of the standard model with additional triplet Higgs, after electroweak symmetry breaking, besides the SM-like Higgs there are 6 additional scalars present in the spectrum which can be denoted as $A^0$, $H^0$, $H^\pm$, $H^{\pm\pm}$ where $A^0$, $H^0$ are the extra CP-odd/even neutral scalars,  $H^\pm, H^{\pm\pm}$ are the charged Higgs and doubly-charged Higgs respectively. The charged Higgs or the doubly-charged Higgs can be pair-produced through the Drell-Yan process, providing good channels to probe the triplet Higgs at the colliders. The ATLAS group already performs a search for the doubly-charged Higgs assuming it mostly decaying into dileptons, and the mass of the doubly-charged Higgs $H^{\pm \pm}$ up to around 800 GeV has been excluded  \cite{CMS:2017pet,ATLAS:2017xqs}\footnote{Recently ATLAS updates their search result and a stronger limit is derived\cite{ATLAS:2022pbd}}. Depending on the number of observed leptons, the detection strategy is classified mainly into three categories: the four-lepton channel, three-lepton channel, and two-lepton channel. Each channel has different sensitivity and the final result is derived from the combination of these three channels. On the other hand, since the triplet Higgs is a triplet under the SM $SU(2)_L$ group, the charged Higgs can be produced together with the doubly-charged Higgs. This production rate can be even higher than the $H^{\pm \pm}$ pair production\cite{Akeroyd:2005gt}. Noticing that the charged Higgs decays into a lepton and a neutrino, the $H^{\pm \pm} H^{\mp}$ production will also contribute to the ATLAS search channels and a better sensitivity could be derived. We will demonstrate this point later.

In addition, since the charged Higgs would decay into a charged lepton and a neutrino, a large missing energy would be present for the $H^{\pm \pm} H^{\mp}$ pair production. It would be intriguing to search for the  $H^{\pm \pm} H^{\mp}$ pair production via the signal of $3e+ {E}^{\rm miss}_{\rm T}$,  which may provide a good sensitivity to the triplet Higgs. This paper is organized as follows: in Sec.~\ref{type2} we give a brief introduction of the type II seesaw model and the mechanism of Type II seesaw leptogenesis. In Sec.~\ref{production} we calculate the production of the $H^{\pm}$ and $H^{\pm \pm}$ at the LHC. We analysis the sensitivity of the triplet Higgs at LHC including the contribution of $H^{\pm \pm} H^{\mp}$ pair production, then we show the prospect for the $H^{\pm \pm} H^{\mp}$ searches requiring a large missing energy for the final states in Sec.~\ref{3l}. We draw our conclusion in Sec.~\ref{conclude}.  

\section{Type II seesaw model}
\label{type2}

The scalar sector of type II seesaw model contains the SM Higgs doublet $\Phi$ and a $SU(2)_{L}$  triplet scalar field $\Delta $ with hypercharge $Y=$ 1  which can be written as

\begin{equation}
\Delta=  \begin{pmatrix}
 \frac{\delta ^{+}}{\sqrt{2}}& \delta ^{++}  \\
 \delta ^{0}&-\frac{\delta ^{+}}{\sqrt{2}}  \\
\end{pmatrix} 
\quad\quad\quad
\Phi= \begin{pmatrix}
 \phi ^{+}\\\phi ^{0}\end{pmatrix}.
\end{equation}
The most general renormalizable and gauge invariant
Lagrangian for the scalar sector is,
\begin{equation}
\mathcal{L} \supset \left ( D_{\mu } \Phi \right )^{\dagger }D^{\mu }\Phi +Tr\left ( D_{\mu  }\Delta  \right )^{\dagger }D^{\mu }\Delta -V\left ( \Phi ,\Delta  \right ).
\end{equation}
Besides SM Yukawa interaction, one can include an additional Yukawa interaction term between the triplet Higgs and leptons,
\begin{equation}
\mathcal{L}_{\nu}=-y_{\nu }L^{T}Ci\sigma _{2}\Delta L+h.c. ,
\end{equation}
where the $y_{\nu }$ is the Yukawa coupling, $L$ is the left-handed lepton doublet and $C$ is the charge conjugation operator. 
After the spontaneous electroweak symmetry breaking (EWSB), the neutral part of $\Delta$ and $\Phi$ acquire a non-vanishing vacuum, \begin{equation}
\left< \Delta \right >= \begin{pmatrix}
0 & 0 \\
\frac{v_{\Delta}}{\sqrt{2}} & 0 \\
\end{pmatrix}
\quad\quad\quad 
\left< \Phi \right >= \begin{pmatrix}
0  \\
\frac{v_{\Phi}}{\sqrt{2}} \\
\end{pmatrix},
\end{equation}
where $v_{\Delta}$ is the vacuum expectation value of the neutral part of triplet Higgs. Then the neutrino mass can be generated by
\begin{equation}
m_{\nu }= \sqrt{2}y_{\nu }v_{\Delta}.
\end{equation}
Here $m_{\nu }$ is a complex symmetric $3 \times 3$ matrix and the physical neutrino masses can be derived by diagonalizing $m_{\nu }$ with PMNS matrix $U$.
The gauge invariant potential for the scalar sector can be written as follows:
\begin{eqnarray}
V(\Phi,\Delta)&=&-m_{\Phi}^2\Phi^\dagger\Phi +   m_{\Delta}^2\rm{Tr}(\Delta^\dagger\Delta)+\left(\mu \Phi^Ti\sigma_2\Delta^\dagger \Phi+\rm{h.c.}\right)+\frac{\lambda}{4}(\Phi^\dagger\Phi)^2 \nonumber\\
&+&\lambda_1(\Phi^\dagger\Phi)\rm{Tr}(\Delta^\dagger\Delta)+\lambda_2\left[\rm{Tr}(\Delta^\dagger\Delta)\right]^2+\lambda_3\rm{Tr}[(\Delta^\dagger\Delta)^2]+\lambda_4\Phi^\dagger\Delta\Delta^\dagger\Phi,
\end{eqnarray}
where $m_{\Phi}^2$ and $m_{\Delta}^2$ are the mass parameters and the $\mu$ term provides a  source of lepton number violation. The $\mu$ term violates lepton number two units for the lepton number assignments of $l_{\Delta }=-2,l_{\Phi }=0$.

After electroweak symmetry breaking we have a state of doubly-charged Higgs $H^{\pm \pm }\left ( \equiv \delta ^{\pm \pm } \right )$, two states of charged scalars $H^{\pm}$ and $G^{\pm}$ which are combinations of $\delta ^{\pm}$ and $\phi ^{\pm}$, and the CP-even neutral states $H^{0}$, $h^{0}$ as well as the CP-odd states $A^{0}$, $G^{0}$, where $G^{\pm}$ and $G^{0}$ are the Goldstone bosons which will ultimately give the longitudinal degrees of freedom of the $W^{\pm}$ and Z bosons.
The mass-squared of the doubly-charged Higgs is given,
\begin{equation}
m_{H^{\pm \pm }}^{2}= \frac{\sqrt{2}\mu v_{\Phi}^{2}-2\lambda _{3} v_{\Delta}^{3}-\lambda _{4} v_{\Phi}^{2} v_{\Delta}}{2 v_{\Delta}}.
\end{equation}
The mass-squared of charged Higgs is
\begin{equation}
m_{H^{\pm  }}^{2}=\frac{2\sqrt{2}\mu v_{\Phi}^{2}+4\sqrt{2}\mu v_{\Delta}^{2}-\lambda _{4} v_{\Delta} v_{\Phi}^{2}-2\lambda _{4}  v_{\Delta}^{3}}{4 v_{\Delta}}.
\end{equation}
For the mass of the CP-even/odd scalars, one can get:
\begin{equation}
m_{H^{0}}^{2}=\frac{1}{2}[A+C+\sqrt{\left ( A-C \right )^{2}+4B^{2}}],
\end{equation}
\begin{equation}
m_{h^{0}}^{2}=\frac{1}{2}[A+C-\sqrt{\left ( A-C \right )^{2}+4B^{2}}],
\end{equation}
\begin{equation}
m_{A^{0}}^{2}=\frac{\mu \left ( v_{\Phi}^{2} +4v_{\Delta}^{2}\right )}{\sqrt{2}v_{\Delta}},
\end{equation}
where
\begin{equation}
A=\frac{\lambda }{2} v_{\Phi}^{2},\quad  B=-\sqrt{2}\mu v_{\Phi}+\left ( \lambda _{1}+\lambda _{4} \right )v_{\Phi} v_{\Delta},\quad  C=\frac{\sqrt{2}\mu v_{\Phi}^{2}+4\left ( \lambda _{1}+\lambda _{4} \right )v_{\Delta}^{3}}{2v_{\Delta}}.
\end{equation}
In the limit of $v_{\Delta} \ll v_{\Phi}$,  we have following masses relation of the physical eigenstates, 
\begin{equation}
m_{H^{\pm \pm }}^{2}-m_{H^{\pm }}^{2}\approx m_{H^{\pm }}^{2}-m_{H^{0}/A^{0}}^{2}\approx -\frac{\lambda _{4}v_{\Phi}^{2}}{4}
\end{equation}
One can define the mass-splitting parameter $\Delta m =m_{H^{\pm \pm }}-m_{H^{\pm}}$ which describes the typical mass difference of the spectra for the triplet Higgs sector. The decay behavior of the triplet Higgs is different from different parameter spaces\cite{PhysRevD.78.015018,Aoki:2011pz,Ashanujjaman:2021txz,Mandal:2022zmy}.For $\Delta m < O(10)$ GeV and $v_{\Delta}< 10^{-4}$ GeV,  the $H^{\pm \pm }/H^{\pm }$ decays into $l^{\pm \pm }/l^{\pm }\nu $. For $\Delta m<  O(10)$ GeV and $v_{\Delta} >  10^{-4}$ GeV,   $H^{\pm \pm }/H^{\pm }$ decays into $W^{\pm \pm }/W^{\pm }Z$ or $W^{\pm }h^{0}$. If $\Delta m >  O(10)$GeV, the cascade decay channels would become significant. In the case of triplet Higgs leptogenesis,  we have $\Delta m < O(5)$ GeV and $v_{\Delta}< 10$ keV~\cite{Han:2022ssz}, thus the $H^{\pm \pm }/H^{\pm }$  would mainly decay into dileptons, giving a typical multi-lepton signature at LHC. In the following we briefly discuss how to achieve leptogenesis.

\subsection{Leptogenesis through type II seesaw}

It is known that the minimal type II seesaw model can not successfully lead to thermal leptogenesis, thus the Affleck-Dine mechanism is considered. In the Affleck-Dine mechanism, the scalar field acquires a large vev along the flat direction during the inflationary epoch. In the subsequent evolution, if the scalar field carries a nonzero baryon or lepton number, the baryon or lepton number violating interactions will induce a rotating trajectory for the vev, which can generate baryon or lepton asymmetry and transfer to ordinary particles at the end of inflation. Fortunately, the Affleck-Dine mechanism can be achieved in the minimal type II seesaw model. 

Considering the non-minimal couplings of $\Delta$ and $\Phi$ to gravity, the relevant Lagrangian in Jordan frame can be written as
\begin{equation}
    \frac{\mathcal{L}}{\sqrt{-g}}\supset -\frac{1}{2}M^2_P R - f(\Phi,\Delta) R + g^{\mu\nu}(D_\mu \Phi)^\dagger(D_\nu \Phi) + g^{\mu\nu}\text{Tr}(D_\mu \Delta)^\dagger(D_\nu \Delta) - V(\Phi,\Delta)
\end{equation}
where $R$ is Ricci scalar. To simplify the analysis, we focus on the neutral components $\phi^0$ and $\delta^0$ and consider the non-minimal coupling to be
\begin{equation}
    f(\Phi,\Delta)=\xi_\Phi |\phi^0|^2 + \xi_\Delta |\delta^0|^2
\end{equation}
Through a Weyl transformation, the Lagrangian can be written in Einstein frame, in which the gravitational portion is of Einstein-Hilbert form. It can be shown that the scalar potential in Einstein frame is
\begin{equation}
    V_E(\Phi,\Delta)=\frac{M_P^4}{\left(M_P^2+2f(\Phi,\Delta)\right)^2}V(\Phi,\Delta)
\end{equation}
which exhibits a flat direction at the large field limit of $\phi^0$ and $\delta^0$. This flat direction can be recognized as a Starobinsky-like inflationary trajectory, and the inflaton is the mixing of $\delta^0$ and $\phi^0$. 

Since the triplet $\Delta$ carry lepton number $l_\Delta=-2$ and the $\mu$ term induces lepton number violating interaction, all the ingredients of Affleck-Dine mechanism are included. During the inflationary evolution, the non-trivial motion of the angular direction of $\delta^0$ can generate lepton number asymmetry, which transfer to ordinary particles during reheating. After reheating, a part of net lepton number is converted to baryon number through the sphaleron process.

However, if any lepton number violating processes are in thermal equilibrium after reheating, the generated lepton asymmetry will be washed out. We require that the processes $L L \leftrightarrow H H$ and $H H \leftrightarrow \Delta$ are never in thermal equilibrium, 
\begin{equation}
\left.\Gamma\right|_{T=m_{\Delta}}=n\langle\sigma v\rangle \approx y^{2} \mu^{2} / m_{\Delta}<\left.H\right|_{T=m_{\Delta}}
\end{equation}
\begin{equation}
\label{eq:washout_condition}
    \left.\Gamma_{I D}(H H \leftrightarrow \Delta)\right|_{T=m_{\Delta}}
    \simeq \frac{\mu^{2}}{32 \pi m_{\Delta}}<\left.H\right|_{T=m_{\Delta}}
\end{equation}
where $\left.H\right|_{T=m_{\Delta}}=\sqrt{\frac{\pi^{2} g_{*}}{90}} \frac{m_{\Delta}^{2}}{M_{p}}$. Using $v_{\Delta} \simeq-\frac{\mu v_{\mathrm{EW}}^{2}}{2 m_{\Delta}^{2}}$ and Eq.~\ref{eq:washout_condition}, the necessary condition to avoid washout effect is found to be
\begin{equation}
    v_{\Delta} \lesssim 10^{-5} \mathrm{GeV}\left(\frac{m_{\Delta}}{1 \mathrm{TeV}}\right)^{-1 / 2}.
\end{equation}
For $m_{\Delta} \gtrsim 1 \mathrm{TeV}$, we require that $v_{\Delta} \lesssim 10 \mathrm{keV}$ to prevent the washout effect and achieve successful leptogenesis.

\section{Production and decay of the triplet higgs}
\label{production}
\begin{figure}[h]
    \centering
    \includegraphics[width=0.4\textwidth]{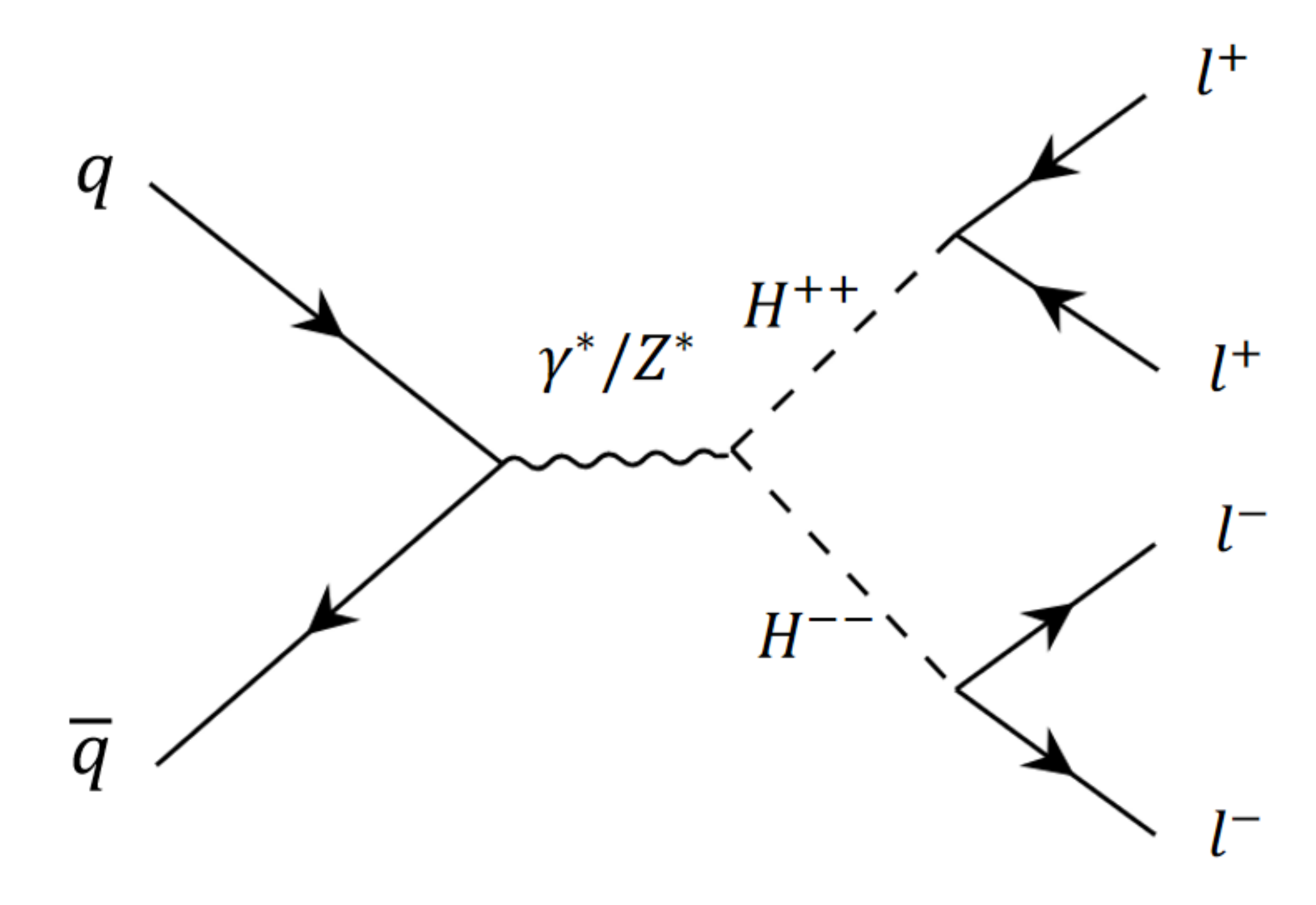}
     \includegraphics[width=0.4\textwidth]{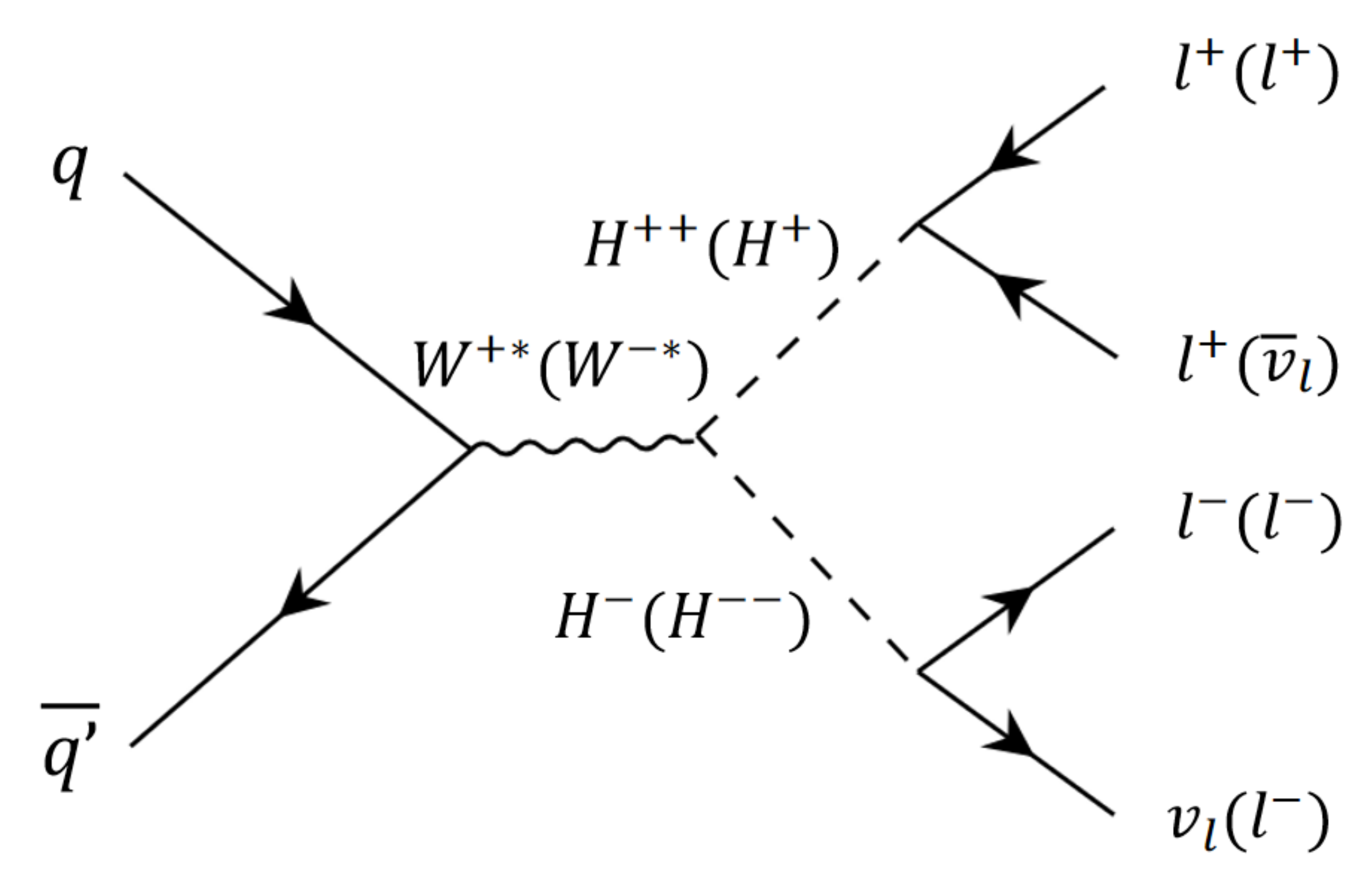}
    \caption{Feynman diagrams of the pair production process $ pp \to H^{\pm \pm }H^{\mp \mp }$ and $ pp \to H^{\pm \pm }H^{\mp }$.}
    \label{fig:feynman diagram}
\end{figure}

The triplet Higgs can be produced at the LHC by the neutral current and charged current Drell-Yan process,
\begin{center}
$q\bar{q} \overset{{\gamma ^{*}/Z^{*}}} {\longrightarrow}H^{\pm \pm }H^{\mp \mp }/H^{\pm }H^{\mp  }/H^{0}A^{0}$
$  \quad \quad \quad q\bar{q}^\prime\overset{W^{*}}{\longrightarrow}H^{\pm \pm }H^{\mp }/H^{\pm }H^{0}/H^{\pm }A^{0}$
\end{center}
and the Feynman diagrams for the $H^{\pm \pm }H^{\mp \mp }$, $H^{\pm \pm }H^{\mp }$ production are presented in Fig.~\ref{fig:feynman diagram}.
\begin{eqnarray}
q(p_1) \, + \, \bar{q}(p_2) \, &\rightarrow &
H^{++}(k_1)\, + \, H^{--}(k_2)\nonumber\\
q(p_1) \, + \, \bar{q}'(p_2) \, &\rightarrow &
H^{++}(k_1)\, + \, H^{-}(k_2)\nonumber
\end{eqnarray}
The parton
level cross section at leading
order (LO) for these processes are:
\begin{eqnarray}
{d\sigma\over dy}(q \bar{q}\rightarrow H^{++}H^{--}) &=& \frac{3\pi \alpha^2 \beta_1^3 (1-y^2)}{N_c {s}}
\Big\{ e_q^2 + 
\frac{ {s}}{({s}-M_Z^2)^2}
\, \frac{\cos 2\theta_W}{\sin^2 2\theta_W}
\nonumber\\
&&\times\Big[4 e_q g_V^q  ({s}-M_Z^2)
 + 4 (g_V^{q2}+g_A^{q2}) {s}\  \frac{\cos 2\theta_W}{\sin^2 2\theta_W}
\Big]
\Big\} ,
\end{eqnarray}
\begin{eqnarray}
{d\sigma\over dy}(q \bar{q}'\rightarrow H^{++}H^{-}) = \frac{\pi \alpha^2 \beta_2^3(1-y^2)}{16 N_c \sin^4\theta_W}\frac{s}{(s-M^2_W)^2},
\end{eqnarray}
where $y= p_1 \cdot k_1$, $s$ is the partonic centre-of-mass energy, and $\alpha$ is the QED coupling evaluated at the scale $\sqrt{s}$. $e_q$ is the electric charge of the quark $q$. $\beta_1= \sqrt{1-4m_{H^{\pm \pm }}^{2}/s}$, $\beta_2= \sqrt{(1-(m_{\pm}+m_{H^{\pm\pm}})^2/s)
(1-(m_{\pm}-m_{H^{\pm\pm}})^2/s)}$. 

\begin{figure}[htbp]
    \centering
    \includegraphics[width=0.8\textwidth]{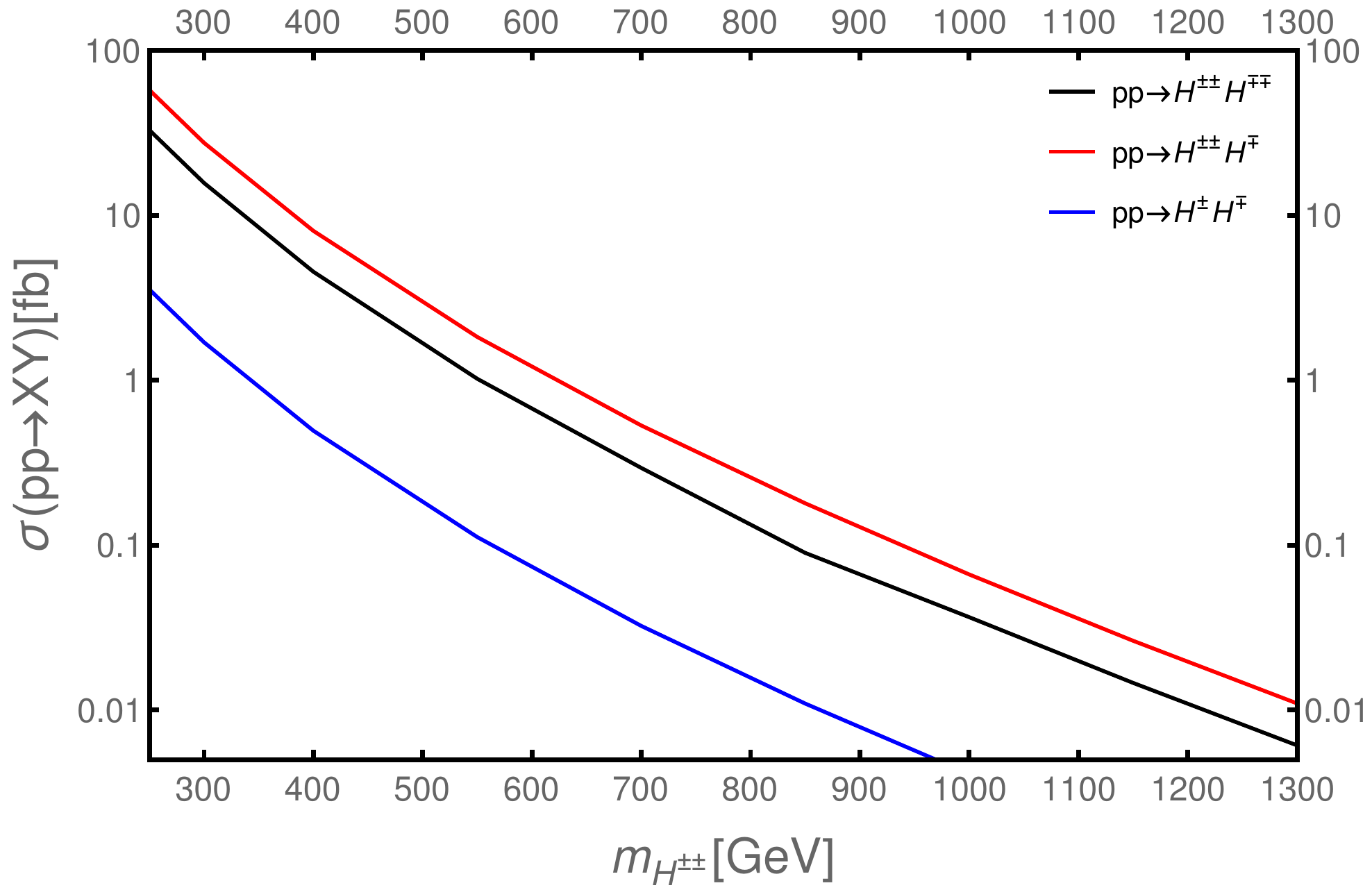}
    \caption{Pair production cross-sections of the triplet scalars at $\sqrt{s} =$ 13 TeV for $\Delta m=0$.}
    \label{fig:cross section}
\end{figure}

In Fig.~\ref{fig:cross section} we show the the cross section of $H^{\pm \pm }H^{\mp \mp }, H^{\pm }H^{\mp }, H^{\pm \pm }H^{\mp }$ pair production with a varying mass of the triplet Higgs. We consider a K-factor as 1.25 \cite{Muhlleitner:2003me} for Fig.~\ref{fig:cross section}, and we assume $H^{\pm \pm }$, $H^{\pm}$ share a same mass parameter.
Doubly-charged triplet Higgs has a considerable cross section and a distinctive decay signature, the same-charge lepton final state. Note that the  $H^{\pm \pm }H^{\mp }$ has an even larger cross section than the $H^{\pm \pm }H^{\mp \mp }$ production, thus it may provide a better sensitivity of the triplet Higgs search.

 The decay modes of triplet Higgs with different $v_{\Delta}-\Delta{m}$ parameters  have been thoroughly discussed in Refs.\cite{Ashanujjaman:2021txz,Mandal:2022zmy,FileviezPerez:2008jbu,Aoki:2011pz}. We consider $\Delta m <  O(1)$GeV and $v_{\Delta}< 10^{-6}$GeV, thus the doubly-charged Higgs $H^{\pm \pm }$ and singly-charged scalars mostly decay to leptonic final states. The decay branching ratio are given by,

\begin{equation}
BR(H^{\pm \pm }\to l_{i}^{\pm }l_{j}^{\pm })= \frac{2}{1+\delta _{ij}}\frac{\left|y_{ij }^{\nu} \right|^{2}}{\sum_{mn}^{}\left|y_{mn }^{\nu} \right|^{2}},
\end{equation}
\begin{equation}
BR(H^{\pm}\to l_{i}^{\pm }\nu _{j})= \frac{\left|y_{ij } \right|^{2}}{\sum_{mn}^{}\left|y_{mn } \right|^{2}},
\end{equation}
with $y_{\nu }=\frac{1}{\sqrt{2}v_{\Delta}}Udiag({m_{1},}m_{2},m_{3})U^{T}$ and $y=\frac{cos\beta }{v_{\Delta}}diag({m_{1},}m_{2},m_{3})U^{T}$, where $U$ is the lepton mixing matrix measured in neutrino oscillation experiments.
 The leptonic branching ratio also depends on the mass order of the neutrino as well as the neutrino mass spectrum. It has been found that for normal hierarchy(NH) and inverted hierarchy(IH)\cite{PhysRevD.78.015018}, 

\begin{center}
NH: $BR(H^{++}\to \mu \mu ),BR(H^{++}\to \tau \tau )\gg BR(H^{++}\to ee)$

IH: $BR(H^{++}\to ee )\gg BR(H^{++}\to \mu \mu  ),BR(H^{++}\to \tau \tau )$
\end{center}
In the following study, we assume the $BR(H^{++}\to ee )=100\% $ to present our result.

\section{Multielectron searches at the LHC}
The ATLAS collaboration has released a multilepton final states search with an integrated luminosity of 36.1$\mathrm{fb^{-1}}$  of pp collisions at $\sqrt{s} =$ 13 TeV \cite{ATLAS:2017xqs}. This analysis focuses on the decays $H^{\pm \pm }\to e^{\pm }e^{\pm },H^{\pm \pm }\to e^{\pm }\mu ^{\pm }$ or $H^{\pm \pm }\to \mu ^{\pm }\mu ^{\pm }$ with a branching ratio around 100\%. The events are divided into three signal regions. We refer to their selection criteria and exhibit in Tab.~\ref{tab:cut}. The final state events with $2,3$ electrons is also considered due to the missing of the electrons in the detector.
In our paper, we first simulated the experimental process by adding the contribution of  $H^{\pm \pm }H^{\mp }$ to the signal event since it will also contribute the $2,3$ electron signal region.
In our simulation, we implement the triplet Higgs model in FeynRules \cite{Alloul:2013bka}, and import UFO files \cite{Degrande:2011ua} into MadGraph \cite{Alwall:2014hca} to generate signal events. We use the NNPDF23LO1 \cite{NNPDF:2014otw} for parton distribution function and the parton showering and hadronisation are simulated with PYTHIA8 \cite{Sjostrand:2014zea}. We perform the detector simulations with Delphes \cite{deFavereau:2013fsa} and data analysis with ROOT \cite{Brun:1997pa}.

For the two-electrons and three-electrons signal regions(SR2E and SR3E), at least one pair of electrons with the same charge is required. The separation of the same-charge electrons and the scalar sum of the electrons' transverse momenta are required to be $\Delta R\left ( e^{\pm }e^{\pm } \right )> 3.5$ and $\sum\left|P_{T}\left ( e \right ) \right| > 300$ GeV, respectively. The vector sum of the electrons’ transverse momenta is required to be $P_{T}\left ( e^{\pm } e^{\pm }\right )> 100$ GeV. The selection criteria for electron are $\left|\eta  \right|< 2.47$ and $P_{T}> 30$ GeV.
Besides the pre-selection cut described above, for the signal region SR3E and SR4E, events are rejected if any opposite-charge same-flavour electron pair is within 10 GeV of the $Z$ boson mass to reduce the background from $Z$ production. In the four-electron signal region(SR4E), there must be two electron pairs with the same charge and the total charge is zero. The $\Delta M/\bar{M}$ requirement is applied to exclude background where the two same-charge pairs have incompatible invariant masses $\left ( \Delta M= \left| m^{++}-m^{--}\right| ,\bar{M}= \frac{m^{++}+m^{--}}{2} \right )$. In the ALTAS experiment, for different $\bar{M}$, the value of $\Delta M$ is different. 
And we simply take $\Delta M/\bar{M}<0.1$ in the four electron channel . In all signal regions, the invariant mass of same-charge electron pairs are required to be above 200 GeV. In order to restrain background events arising from top-quark decays, events with b-tagged jet are vetoed.

\begin{figure}[htbp]
    \centering
    \includegraphics[width=0.8\textwidth]{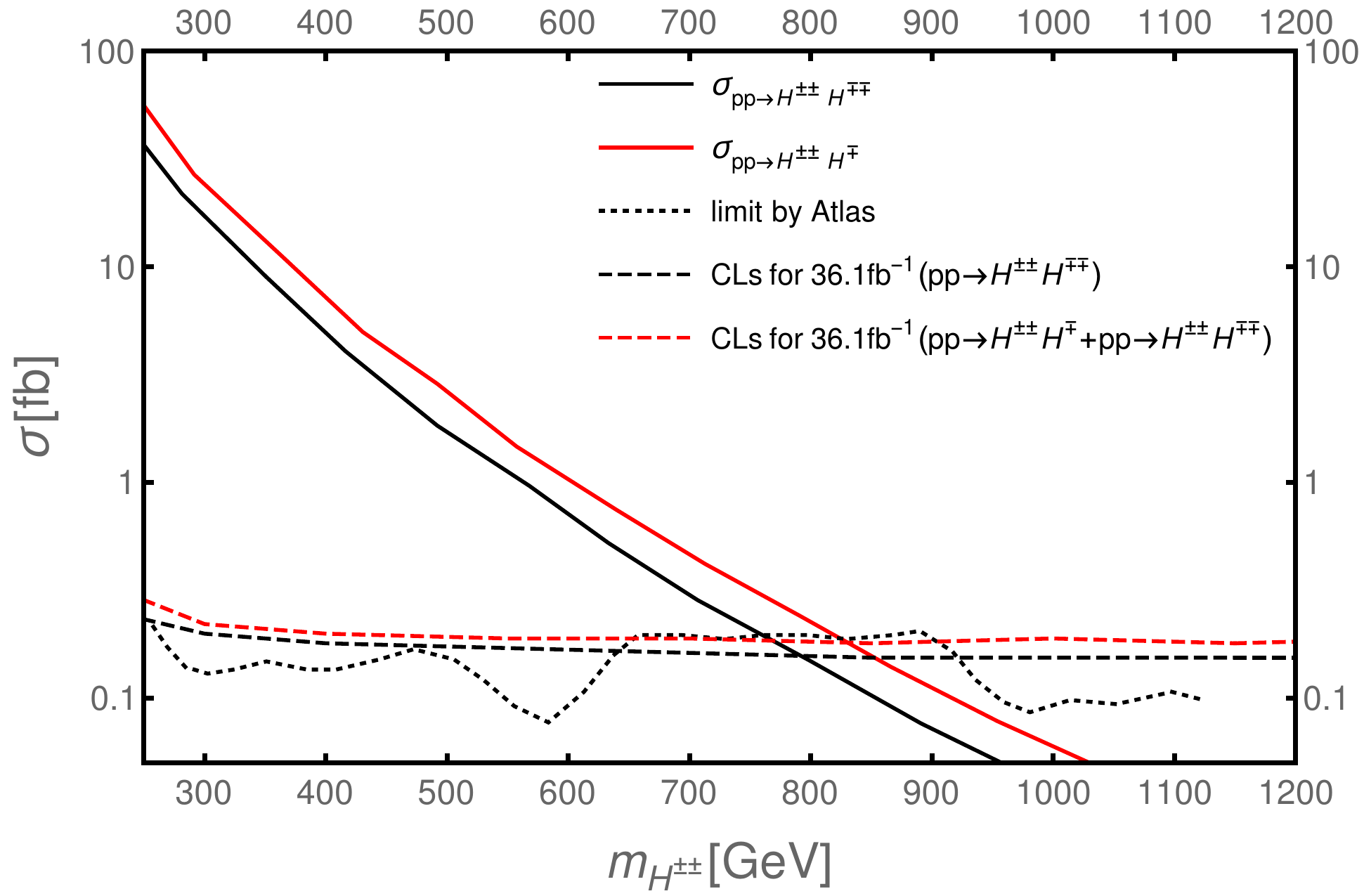}
    \caption{The limits for $B(ee)/B(e\mu )/B(\mu \mu )=100\%/0\%/0\%$. The black and red solid lines represent the production cross section of the $pp\to H^{\pm \pm }H^{\mp \mp }$ and $pp\to H^{\pm \pm }H^{\mp }$ process. The black dashed line is the 95$\%$ CL limit we get for the $pp\to H^{\pm \pm }H^{\mp \mp }$ process, which is comparable to the limit obtained by Atlas and depicted as black 
   dotted line. The red dashed line is the 95$\%$ CL limit we get by adding the contribution of the two processes together.}
    \label{fig:limit}
\end{figure}

\begin{table}[h]

\centering
  	
  	\begin{tabular}{|c|c|c|c|}
  		\hline
  	                        & SR2E  & SR3E  &  SR4E          \\ 
  		\hline
  		b-jet veto           & $\circ $   &  $\circ $   &   $\circ $          \\
  		\hline
  		$Z$ veto           &    &  $\circ $    &   $\circ $            \\
  		\hline
  		$P_{T}\left ( e^{\pm } e^{\pm }\right )> 100$GeV           & $\circ $    &  $\circ $    &              \\
  		\hline
  		$\sum\left|P_{T}\left ( e \right ) \right| > 300$GeV           & $\circ $    & $\circ $     &              \\
  		\hline
  		$\Delta R\left ( e^{\pm },e^{\pm } \right )< 3.5$           & $\circ $    &  $\circ $    &               \\
  		\hline
  		$\Delta M/\bar{M} $           &     &      &   $\circ $            \\
  		\hline

  	\end{tabular}
        \caption{ Selection criteria in all the signal region }
  	\label{tab:cut}	
     \end{table}
%
To validate our simulation, we first simulate the signal events from $pp\to H^{\pm \pm }H^{\mp \mp }$ production and get the signal cut efficiency. Using the observed signal event from the article, we apply the $CLs$ method \cite{Junk:1999kv} to get the 95$\%$ CL upper limits on the $pp\to H^{\pm \pm }H^{\mp \mp }$ cross section. The result is shown in Fig.~\ref{fig:limit} denoted as the black dashed curve \cite{ATLAS:2017xqs}. As a comparison, the limit from ATLAS experiments are also shown as the black dotted line.  It shows our limit is close to the one derived from the ATLAS experiment.   

Since $pp\to H^{\pm \pm }H^{\mp }$ contributes the SR2E and SR3E signal region,  we expect the real limit should be stronger. Therefore, we simulate the process $pp\to H^{\pm \pm }H^{\mp }\to l^{\pm }l^{\pm }l^{\mp }\nu ^{\pm }$ and get the corresponding signal efficiency. To combine our result, we denote the $\sigma _{1,2}$ and $\varepsilon_{1,2}$ as cross-section and cut efficiency for $pp\to H^{\pm \pm }H^{\mp \mp}$ process and $pp\to H^{\pm \pm }H^{\mp }$ process respectively, then the total signal events  
    $n=\mathcal{L}\sigma _{1}\varepsilon _{1}+\mathcal{L}\sigma _{2}\varepsilon _{2}$ for each signal region. We set the limit on the total signal events. To show our result, we can use an effective cut efficiency of $\varepsilon_{2eff}=\varepsilon _{2} + \sigma_1 /\sigma _{2} \varepsilon _{1}$ for  $pp\to H^{\pm \pm }H^{\mp }$ process and set the limit on the cross section of $pp\to H^{\pm \pm }H^{\mp }$ production, which is shown as the red dashed curve in Fig.~\ref{fig:limit}.  It shows the combined limit is around 100 GeV stronger than the one derived only from $pp\to H^{\pm \pm }H^{\mp \mp }$ process.


\section{$3e+ {E}^{\rm miss}_{\rm T}$ signal}
\label{3l}
Notice that $pp\to H^{\pm \pm }H^{\mp }$ has a relatively larger cross section and the final states include a missing energy. It is intriguing to examine whether the $3e+ {E}^{\rm miss}_{\rm T}$ could provide a better sensitivity to the triplet Higgs. 

The relevant background for this signal mainly originates from diboson$(ZZ,ZW,WW)$, $t\bar{t}$, $t\bar{t}W$, $t\bar{t}Z$, $t\bar{t}h$, triboson and Drell-Yan processes. However, as shown in the ATLAS paper, for the $3l$ process, the diboson background is much more dominant than other backgrounds. 
Therefore for the background simulation, we only consider the events from the diboson process. The background and the signal are both simulated by using MadGraph with an MLM matching. For the cross section of the diboson, we also add the K-factor to include the NLO correction. The LO cross-section for the diboson process and the corresponding K-factor at $\sqrt{s} =$ 13 TeV LHC \cite{Campbell:2011bn} are shown in Tab.~\ref{tab:cross}.






\begin{table}[h]
\centering
  	\begin{tabular}{|c|c|c|c|c|}
  		\hline
  	                        & $ZZ$   & $W^{+}Z$  &  $W^{-}Z$  &   $WW$         \\ 
  		\hline
  		 $\sigma_{LO} $[pb] &  9.89  &  15.51    &   9.53     &   67.74        \\ 
  		\hline
  		K-factor           & 1.62    &  1.84     &   1.91     &   1.66         \\
  		\hline
  		
  	\end{tabular}
  	\caption{ The LO cross-sections and K-factors for diboson production at $\sqrt{s} =$ 13 TeV }
  	\label{tab:cross}
  \end{table}

\begin{figure}[h]
    \centering
 	\includegraphics[width=0.55\linewidth]{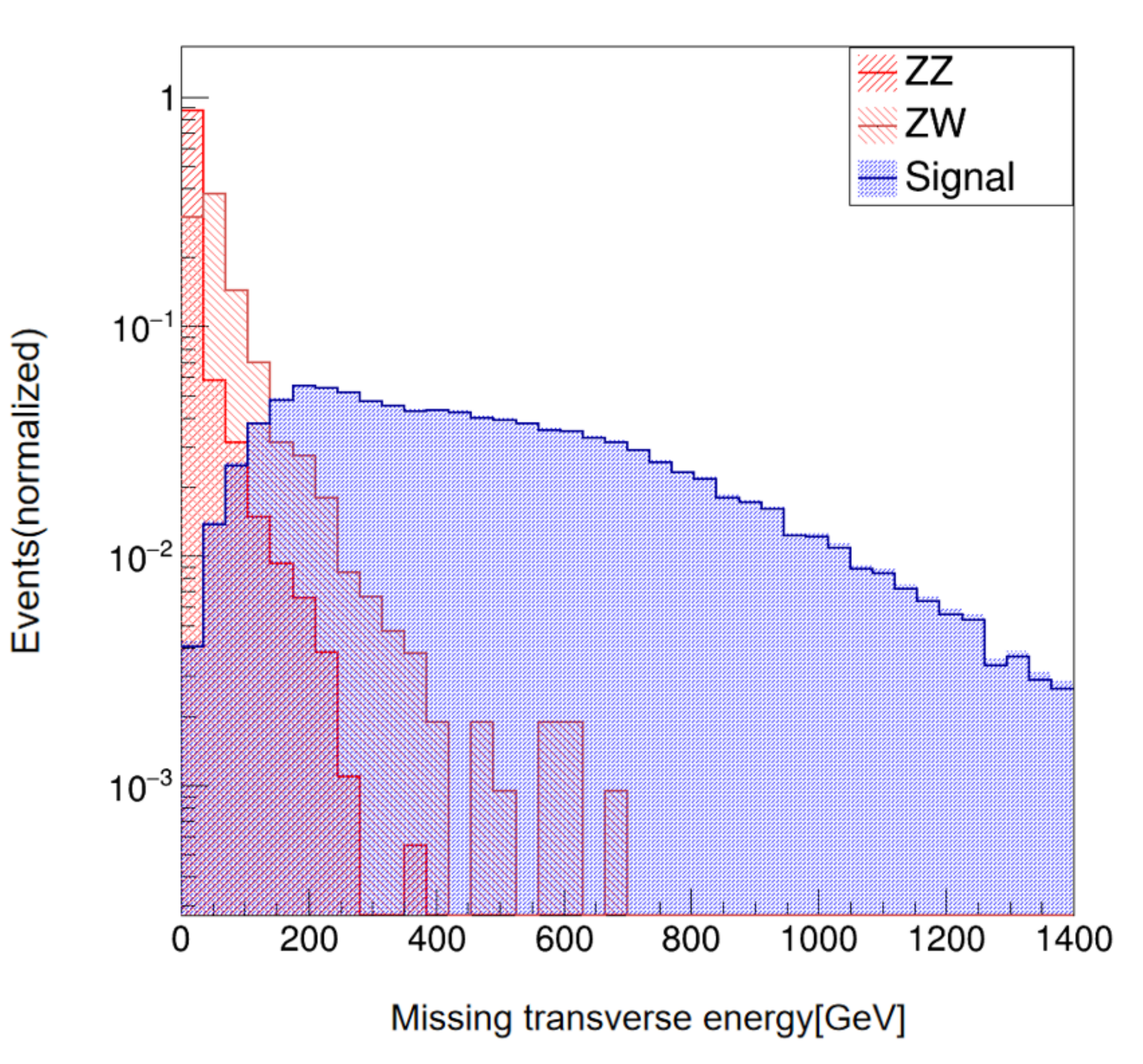}
 	\caption{The missing transverse energy distribution of $pp\to H^{\pm \pm }H^{\mp }\to l^{\pm }l^{\pm }l^{\mp }\nu ^{\pm }$ ($B(ee)=100\%$)process and diboson background with the pre-selection. The mass of $H^{\pm \pm }$, $H^{\mp }$ are assumed to be 1 TeV here. }
 	\label{fig:signal}
\end{figure}

To ensure simulation credibility and validate the charge misidentification effect in the electron channel, the same-charge region(SCR) is also considered, which only exert b-jet veto. For $pp\to H^{\pm \pm }H^{\mp }\to l^{\pm }l^{\pm }l^{\mp }\nu ^{\pm }$, large missing transverse energy will appear in the final states. We show the missing energy distribution of diboson process and $pp\to H^{\pm \pm }H^{\mp }\to l^{\pm }l^{\pm }l^{\mp }\nu ^{\pm }$ process in figure  \ref{fig:signal}. It shows that a cut on the missing energy around few hundred GeV would remove much of the background.

\begin{figure}[h]
    \centering
 	\includegraphics[width=0.8\textwidth]{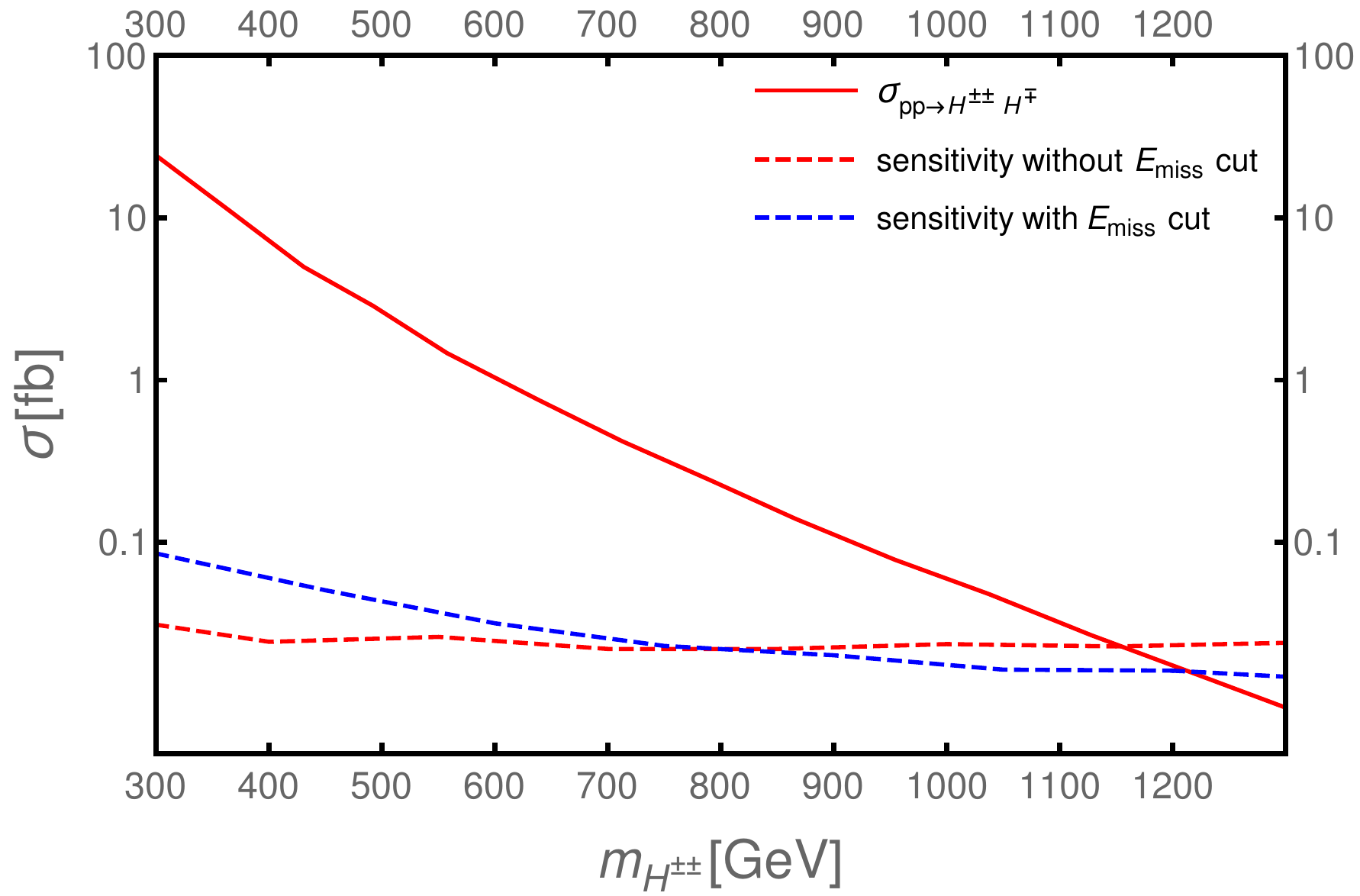}
 	\caption{The sensitivity of future searches with $3000 fb^{-1}$.}
 	\label{fig:sensitivity}
\end{figure}

The distinction of the missing energy distribution between signal and diboson background motivates us to add a missing energy cut $E^{\rm miss}_{\rm T} > 300$ GeV. The cut flow for the background and signal for a luminosity $3000 fb^{-1}$ at 13 TeV LHC are shown in Tab.~\ref{tab:cut flow}. It clearly shows that only 10\% of the background are left after imposing the cut $E^{\rm miss}_{\rm T} > 300$ GeV, while most of the signal events are still kept. Using expected discovery significance $S/\sqrt{B}$, the results are shown in Fig \ref{fig:sensitivity} at 13 TeV LHC with a luminosity $3000 fb^{-1}$. We find the triplet mass less than 1.2 TeV can be reached at $2\sigma$ for the high luminosity LHC in future. As a comparision, we also show the $2\sigma$ sensitivity for the multi-electron searches channels mentioned in last section where the  missing energy cut is not imposed. We find that at when the triplet Higgs mass is below 800 GeV, the multi-electron channel still provide a better sensitivity for the triplet Higgs. However, the $3e+ {E}^{\rm miss}_{\rm T}$ signal could reach a higher triplet Higgs mass when the triplet Higgs mass is larger than 800 GeV. The main reason for this is that when the mass of the triplet Higgs is low, the missing energy could be lower and the missing energy cut would also hurt the signal. We believe that an even larger missing energy cut could further suppress the background and a better sensitivity for the heavy triplet Higgs can be reached.

\begin{table}[h]
\centering
  	
  	\begin{tabular}{|c|c|c|c|c|}
  		\hline
  	                        & Diboson BKG  &   600 GeV   &  900 GeV   &   1200 GeV             \\ 
  		\hline
  		pre-selection     &  14518  &    2249     & 242 &  38         \\
  		\hline
  		$m_{invariant}> 200$GeV    &  3037  &   2199    & 241 &   38         \\
  		\hline
  		$P_{T}\left ( e^{\pm } e^{\pm }\right )> 100$GeV           &  1379  &  2168    &  239  &   37            \\
  		\hline
  		$\sum\left|P_{T}\left ( e \right ) \right| > 300$GeV          &  673  &  2139   &  237 &  37               \\
  		\hline
  		$\Delta R\left ( e^{\pm },e^{\pm } \right )< 3.5$       &  490   &   1596    & 174 &  26               \\
        \hline
  		$E^{\rm miss}_{\rm T} > 300$ GeV & 49.1  & 790   & 111 &   20               \\
  		\hline
  		Significance  &   -  &  113    & 15.8 &    2.9            \\
  		\hline
  		
  	\end{tabular}
  	\caption{The cut flow for the Diboson background and the signal($m_{H^{\pm \pm } }=600$ GeV,$m_{H^{\pm \pm } }=900$ GeV,$m_{H^{\pm \pm } }=1200$ GeV) with an integrated luminosity
of $3000 fb^{-1}$ and $\sqrt{s} =$ 13 TeV.}
  	\label{tab:cut flow}
  \end{table}

\section{Conclusion}
\label{conclude}
The type II seesaw leptogenesis simultaneously explains the origin of neutrino masses, the baryon asymmetry of our universe, and the inflation. The large hadron collider(LHC) provides an opportunity to directly test the type II seesaw leptogenesis by looking for the predicted triplet Higgs. In this paper, we perform an analysis of the detection prospect for the triplet Higgs at the LHC through the multi-electron channels. We find that due to the contribution of $pp\to H^{\pm \pm }H^{\mp }$ process,  the sensitivity of multi-electron channels searching for the doubly-charged Higgs pair production can be improved. We also investigate the $3e+ {E}^{\rm miss}_{\rm T}$ signals to probe the $pp\to H^{\pm \pm }H^{\mp }$ production and we find this channel may provide a better sensitivity than the multi-electron channel. Our result shows that the future LHC could probe a triplet Higgs around 1.2 TeV at 2$\sigma$ level with a luminosity 3000 $fb^{-1}$ for the $3e+ {E}^{\rm miss}_{\rm T}$ search channel.

\acknowledgments

 C. H. is supported by the Sun Yat-Sen University Science Foundation, and by the Fundamental Research Funds for the Central Universities, Sun Yat-sen University under Grant No. 23qnpy58.

\appendix
\section{The $CL_{s}$ method}

Indistinguishable from background hypotheses in the case of few signal events, we use the $CL_{s}$ method to improve experimental sensitivity. The usual confidence level for signal and background hypothesis is given by the probability that the test-statistic $Q$ is less than or equal to the value observed in the experiment: 

\begin{equation}
CL_{s+b}=P_{s+b}(Q\leq Q_{obs})=\int_{-\infty}^{Q_{obs}}\frac{dP_{s+b}}{dQ}dQ,
\end{equation}
where $\frac{dP_{s+b}}{dQ}$ is the probability distribution function for signal and background experiments. 
Likewise, the confidence level in the background-only hypothesis is:

\begin{equation}
CL_{b}=P_{b}(Q\leq Q_{obs})=\int_{-\infty}^{Q_{obs}}\frac{dP_{b}}{dQ}dQ,
\end{equation}
and $\frac{dP_{b}}{dQ}$ is the probability distribution function for background-only experiments. 

To obtain the limit, we use the definition of  $CL_s$

\begin{equation}
CL_{s}=\frac{CL_{s+b}}{CL_{b}},
\end{equation}
The signal hypotheses is excluded at the confidence level $CL$ when    
\begin{equation}
1-CL_{s}\leq CL.
\end{equation}
To combine the results of the signals from several channels, the test statistic is defined as the likelihood ratio 
\begin{equation}
Q= \prod_{i=1}^{n}q_{i}
\end{equation}
with
\begin{equation}
q_{i}=\frac{\frac{e^{-(s_{i}+b_{i})}((s_{i}+b_{i})^{N_{i}}}{N!}}{\frac{e^{-b_{i}}b_{i}^{N_{i}}}{N!}}
\end{equation}
for counting experiments. The estimated signal and background are $s_{i}$ and $b_{i}$, and $i$ labeled the channel. Then $N$ is the number of observed candidates. The final likelihood function should also include the uncertainty of the backgrounds. All of the above calculation can be preformed numerically by Monte Carlo method.

\bibliographystyle{JHEP}
\bibliography{references}

\end{document}